\documentclass[twocolumn]{article}

\usepackage{natbib,natbibspacing}
\usepackage[margin=1.5cm,bottom=2cm,top=2cm]{geometry}
\usepackage{amsmath}
\usepackage{graphicx, tabularx, colortbl}
\usepackage[colorlinks=true,allcolors=blue,urlcolor=magenta,bookmarks=false,pdfstartview=FitV]{hyperref}
\usepackage[ansinew]{inputenc}
\usepackage[english]{babel}
\usepackage{sectsty}
\usepackage{fancyhdr}
\usepackage[small,bf]{caption}
\usepackage{balance}

\sectionfont{\large}
\subsectionfont{\normalsize}

\def\mms{$\mathrm{mm\,s^{-1}}$}
\def\cms{$\mathrm{cm\,s^{-1}}$}
\def\ms{$\mathrm{m\,s^{-1}}$}
\def\mum{$\mu$m}

\definecolor{tablegray}{gray}{0.9}
\newcommand\blfootnote[1]{%
  \begingroup
  \renewcommand\thefootnote{}\footnote{#1}%
  \addtocounter{footnote}{-1}%
  \endgroup
}

\begin{document}

\twocolumn[\begin{@twocolumnfalse}
\title{Normal Collisions of Spheres:\\A Literature Survey on Available Experiments}

\author{C. G{\"u}ttler$^{1,2}$, D. Hei{\ss}elmann$^2$, J. Blum$^2$, S. Krijt$^3$\\
\small $^1$ Department of Earth and Planetary Sciences, Kobe University, 1-1 Rokkodai-cho, Nada-ku, Kobe 657-8501, Japan\\[-0.3em]
\small $^2$ Institut f{\"u}r Geophysik und extraterrestrische Physik, Technische Universit{\"a}t Braunschweig, Mendelssohnstr. 3,\\[-0.3em]
\small D-38106 Braunschweig, Germany\\[-0.3em]
\small $^3$ Leiden Observatory, Leiden University, P.O. Box 9513, 2300 RA Leiden, The Netherlands}
\date{\today}

\maketitle


\begin{abstract} \noindent
The central collision between two solid spheres or the normal collision between a sphere and a plate are important to understand in detail before studying more complex particle interactions. Models exist to describe this basic problem but are not always consistent with available experiments. An interesting benchmark to compare models and experiments is the relation between the normal coefficient of restitution $e$ and the incident velocity $v_\mathrm{i}$. In order to draw a broad comparison between experiments and models (Krijt, S., G{\"u}ttler, C., Hei{\ss}elmann, D., Tielens, A.G.G.M., Dominik, C., Energy dissipation in head-on collisions of spheres, submitted), we provide in this article an overview on the literature describing experiments on normal collisions, preferably providing data on $e(v_\mathrm{i})$. We will briefly summarize our expectation on this relation according to an established collision model in order to classify these experiments. We will then provide an overview on experimental techniques, which we found in the summarized articles, as well as a listing of all experiments along with a description of the main features of these. The raw data on $e(v_\mathrm{i})$ of the listed experiments were digitized and are provided with this article.
\end{abstract}
~\\
\end{@twocolumnfalse}]

%
%
%
%
%
%

\section{Introduction} \label{sec:introduction} \blfootnote{This article was published on the arXiv motivated by our need for a broad survey on the experimental literature as well as the raw data from these. This will be used in the paper of \citet{KrijtEtal:preprint_b}. A future publication of the article at hand in a refereed journal is not excluded but also not foreseen at the moment. In any case, constructive comments are very welcome.}

The problem of dropping a solid sphere from a certain height onto a massive plate of the same material and measuring the rebound velocity is such a fundamental process that it is expected to be understood in detail. Indeed, much work has been done on this problem, which is mandatory for understanding more complex collisional processes which are necessary to understand complex systems related to astrophysics (dust coagulation, Saturn's rings, rubble pile asteroids) or industrial processes (handling of powders, filling of silos). But still, it is not fully understood and there are several striking disagreements between experiments and models.

This study is motivated by recent developments in the field of planet formation where our understanding of collisions of small aggregates of silicate dust -- which should ideally stick to each other and grow into planets -- relies on a good interaction model between the microscopic dust grains. A misfit between laboratory collision experiments and numerical simulations of these \citep{GuettlerEtal:2010, WadaEtal:2011} made us look into the interaction model again to find the reason for this. Moreover, recent attempts to describe collisions of millimeter sized, porous dust aggregates on a macroscopic scale with an established collision model for solid particles \citep{ThorntonNing:1998} turned out to be surprisingly successful \citep{WeidlingEtal:2012} so that any development in our understanding of this simple problem of a sphere dropping onto a plate would stimulate the field of planet formation.

The elastic contact between two spheres or between a sphere and a plate has been described by \citet{Hertz:1881}, \citet{JohnsonEtal:1971} and \citet{Derjaguinetal:1975} added the attractive adhesion between the two surfaces in contact. A conclusive description of a collision including adhesion, elastic, and plastic deformation has been given by \citet{ThorntonNing:1998}. Although there are other models \citep[e.g., for viscous dissipation,][]{BrilliantovEtal:2007} we will restrict our discussion to the model of \citeauthor{ThorntonNing:1998}, which includes an analytic description of the coefficient of restitution $e$ (the ratio of the velocities after and before the collision) as a function of the incident velocity $v_\mathrm{i}$.

Focusing on this relation between the coefficient of restitution and the incident velocity, our attempt is to give an overview on the broad literature on experiments that present data on this relation and, in a second step presented by \citet{KrijtEtal:preprint_a, KrijtEtal:preprint_b}, draw a comparison between those experiments and theoretical models. In earlier works, models have been developed and described but rarely been tested with good experiments -- partly because those were not available -- and if so only to a limited number. A positive example however where this has been done is by \citet{KimDunn:2007}, who compared six different experiments on normal collisions to two different (microsphere) impact models. However, as we will show, many more experiments exist and we believe that it is worth to draw a broader comparison.

\begin{figure}[t]
    \begin{center}
        \includegraphics[width=\columnwidth]{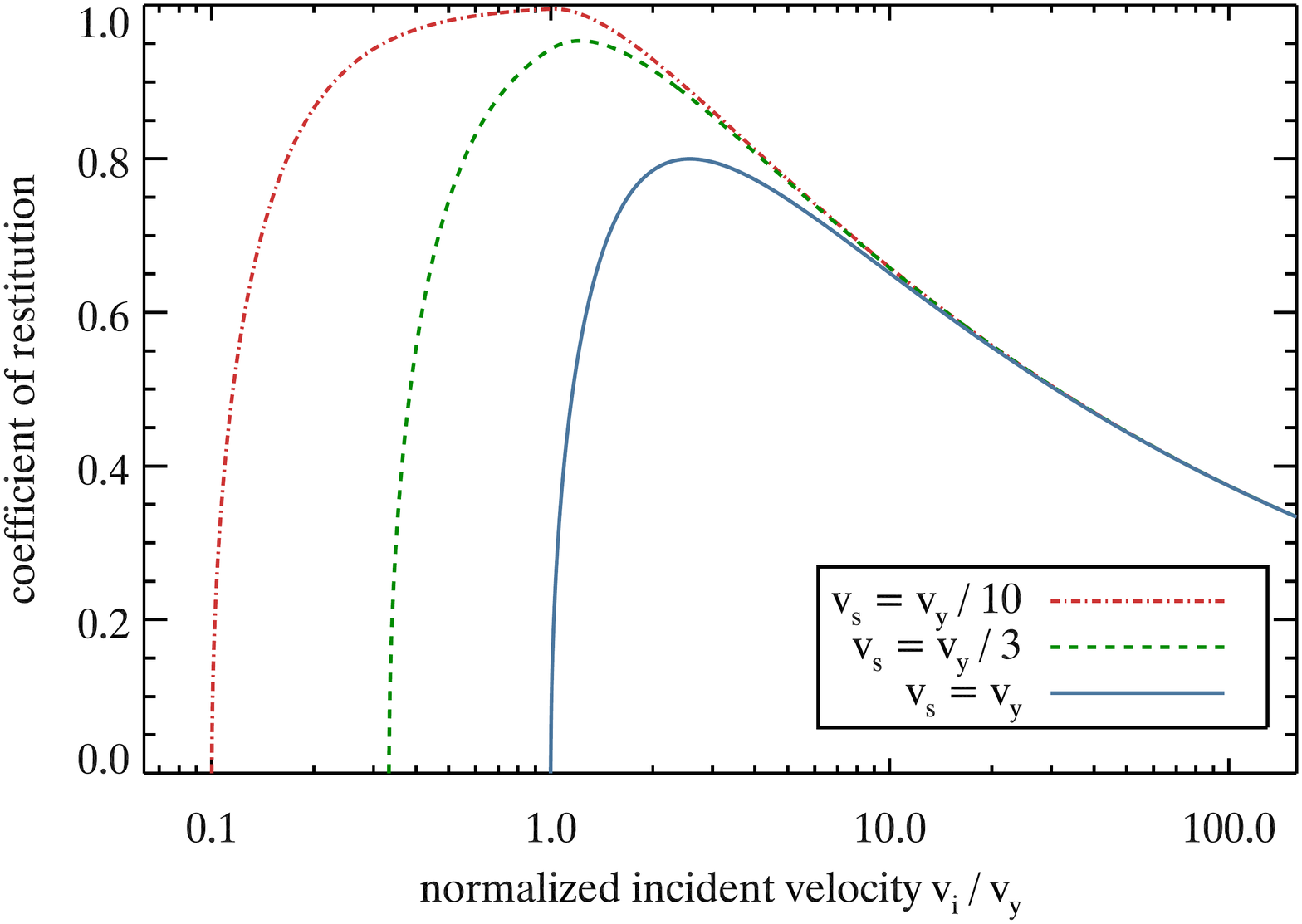}
        \caption{\label{fig:TN_plot}Expected relation for the coefficient of restitution against the incident velocity \citep[after][]{ThorntonNing:1998}.}
    \end{center}
\end{figure}

With a focus on the experimental data of $e(v_\mathrm{i})$, we have the potential to gain information on the adhesive interaction at low velocities as well as the plastic or viscous dissipation at larger velocities. The relation that was developed by \citet[their Eqs. (79) to (81)\footnote{Please note than Eq. (81) of \citet{ThorntonNing:1998} contains a typo, the leading constant should read $6 \sqrt{3}/5$.}]{ThorntonNing:1998} is presented in Fig. \ref{fig:TN_plot}. The curve is parametrized by two velocities: the sticking velocity $v_\mathrm{s}$ is the critical velocity below which all collisions lead to adhesive sticking and the coefficient of restitution is zero. Apart from the size, mass, and elastic properties (Young's modulus, Poisson's ratio) of the colliding particles, the sticking velocity is determined by the surface energy, a sink to dissipate kinetic energy. The second parameter, the plastic yield velocity $v_\mathrm{y}$, is the threshold above which the particles do not behave fully elastic any more. It involves the yield pressure above which the material plastically yields \citep{Johnson:1985}.

The curves in Fig. \ref{fig:TN_plot} show the following characteristics: the coefficient of restitution is zero for velocities smaller than the sticking velocity and then steeply rises as $e = \{ 1 - ( v_\mathrm{s} / v_\mathrm{i} )^2\}^{1/2}$. If the yield velocity is much larger than the sticking velocity (red dash-dotted curve), the coefficient of restitution shows a plateau close to unity for velocities between the stick and yield velocity. If those two parameters are not too different (e.g., blue solid curve), the maximum is well below unity as there is always considerable dissipation of energy, either by the adhesive contact or by plastic deformation. It is very conceivable that some materials may have $v_\mathrm{y} < v_\mathrm{s}$ so that the maximum can be at a relatively small level. The slope of the decline of the coefficient of restitution for velocities larger than the yield velocity and much larger than the stick velocity then approaches $e \propto v_\mathrm{i}^{-1/4}$. It should be noted that the shape of the curve is fixed by the model of \citeauthor{ThorntonNing:1998} but it can be shifted horizontally depending on the yield velocity ($v_\mathrm{i}$ is normalized by $v_\mathrm{y}$ in Fig. \ref{fig:TN_plot}). The maximum, and especially the shape of the maximum is determined by the ratio between stick and yield velocity. If the model is correct, a rather constant coefficient of restitution should only exist near unity, which is not observed in many experiments as we will show later.

Figure \ref{fig:TN_plot} is given here only for illustrative purposes to develop a certain expectation on how the data of the experiments presented below should qualitatively look like to be comparable to established collision models. A detailed comparison between laboratory data and theoretical collision models will be drawn by \citet{KrijtEtal:preprint_b} and our task is now to give an overview on the literature on available experiments. We found 25 experiments (in a larger number of publications) with useful data on $e(v_\mathrm{i})$ which are presented in Table \ref{tab:overview_table}. The data from these publications was digitized and provided with this publication (on the arXiv server). An unbiased choice of this will then be used by \citeauthor{KrijtEtal:preprint_b} to be compared to collision models. In Sect. \ref{sec:methods} we will give a general overview of experimental techniques used to study the normal collisions between two spheres or between a sphere and a plate. In Sect. \ref{sec:review} we will then describe all the experiments provided in Table \ref{tab:overview_table} and refer to the methods used. A first qualitative comparison between some of these experiments and the expectations from Fig. \ref{fig:TN_plot} will then be given in Sect. \ref{sec:discussion} along with a discussion of the surface conditions of the samples used. Surface conditions will turn out to potentially be critical as further discussed by \citeauthor{KrijtEtal:preprint_b} Finally, we hope that this review on the experiments as well as the raw data provided will stimulate further work on the comparison of theory and experiments, to the development of more realistic theories and more precise experiments.

\section{Methods used to study normal collisions}\label{sec:methods}

A detailed description of the experiments in the literature will be given in the next section. Before that, we want to introduce general experimental methods to study normal collisions, which will then be referred to.

\subsection{General techniques}

The most simple experiment one could think of would be to drop a sphere onto a plate and with the help of gravity the collision velocity is then determined by the drop height only. Observing multiple bouncing collisions and a thereby resulting slowdown of the incident velocities, one can also study collisions at velocities corresponding to drop heights which might otherwise not be easily achievable, i.e., fractions of a millimeter and thus velocities less than 0.1 \ms.

Collisions between two particles can be achieved by releasing two particles on top of each other. The upper particle is released first and catches up with the second particle which is released slightly later. The time difference $\Delta t$ determines the relative velocity $v = g_0 \Delta t$ ($g_0=9.81\ \mathrm{m\,s^{-2}}$) of the entirely free collision and with the initial distance one can control the vertical position where the collision takes place. Due to restrictions in the smallest initial distance of the two particles, very small velocities (e.g., 0.01 \ms) can only be achieved with long drop heights.

Higher velocities were achieved when inserting the spheres into two opposite guide tubes, where they are held at one end by an underpressure; the other end is open. The particles are then pushed out by applying an overpressure where the vacuum was before so that the spheres are pushed out in the fashion of an air gun. Velocities up to 60 \ms\ without rotation of the particles were reported with this method.

Another method to study two-particle collision at small velocities is to suspend them on a pendulum, preferably supported by two strings each particle, and release one or both spheres from a certain height. Again, this allows for the study of multiple bouncing collisions and extremely low velocities have been reported, but it should be noted that these collisions are never entirely free and the particles lack some degree of freedom. This does not necessarily have to be a problem but it should be considered. A torsion pendulum was also reported for some experiments with one sphere mounted to the pendulum and another sphere or block steadily fixed. An advantage of pendulum experiments in general is that -- due to simple technical reasons -- the impact parameter can be more precisely tuned than in free collisions.

In microgravity experiments, it is possible to study a system of particles, i.e., a granular gas, which collisionally cools due to the ongoing inelastic collisions. By this, completely free and extremely slow velocities can be observed after some time. For a limited microgravity time this requires a good compromise for the particle density -- dense for fast cooling, sparse for good observability. With this design it is however not possible to control the impact parameter and the rotation of the particles, which are therefore both arbitrary.

\subsection{Particle release}

The release of macroscopic particles can be achieved in different ways: a simple mechanism for magnetic particles is to hold them with an electromagnet which is then switched off. To avoid magnetic fields, the particles can also be held with a small hole connected to a vacuum, which is then instantly flooded to release the particle. Solenoid magnets have been used, for instance in the design of a rotational solenoid magnet. With an initially vertical aligned arm and a spoon-like mount at the end, the particle is released when the spoon is instantly moved down as the arm is rotated around the horizontal axis. Two linear solenoid magnets can be used to hold a particle between them and then synchronously be pulled apart. Also, a linear solenoid can be used to hold a thin string, which is glued to or frozen into the particle and then release it (or oppositely cutting a string). This string must of course be of insignificant mass compared to a much larger particle. Most of these techniques can and have been designed in a way to minimize unwanted rotation.

\subsection{Velocity measurement}

The precise measurement of the velocity before as well as after the collision is crucial for a good measurement of the coefficient of restitution and also here we find many different techniques. The most intuitive technique is probably to use a video camera, preferably one with a high frame rate. The velocities can then directly be determined from the images and if a beam splitter or two cameras are used, one can also determine the 3D velocity vectors in case of non-perfect central collisions. If not available, a stroboscopic flashlight and a photo camera with long exposure can substitute the high-speed camera. Even a combination of a medium fast camera and a stroboscopic flashlight (a pulsed laser in that case) of known and constant illumination duration have been described. The particles then appear on the images as long streaks and the lengths as well as the distance of the streaks yields the velocities.

Doppler systems have been reported for the velocity measurement of micrometer grains. Slightly differing in the details, the common characteristic of these systems is that they produce a Doppler shifted light signal, where the shift is proportional to the particle velocity. This technique is mainly used for droplets or aerosols.

Two successive light barriers with exactly known distance can be used to measure the time difference between the two signals when the particle crosses one of the barriers.

In the simple setup with a spherical particle repeatedly bouncing on a horizontal plate, the velocity can be deduced from the time difference between two bouncers and assuming a perfect parabola. Air friction and other effects must therefore not play a role in those cases. The time can be measured by fixing an acoustic or acceleration sensor to the target plate or even using a sensitive target (e.g., a piezoelectric crystal). A light barrier close to the target would also work.

\subsection{Treating micrometer particles}

While some of the techniques mentioned above can also be applied to micrometer-sized particles, these normally require a special treatment. A major complication with small particles is their tendency to aggregate while one rather wants to study collisions with individual particles. One solution is to freshly form these particles in an aerosol generator (e.g., vibrating orifice) and not to give them a chance to aggregate.

If they are aggregated however, particle dispersers exist, using electrostatic effects so that the particles must be discharged after the disaggregation. Another method is to mechanically disaggregate them in the collisions with the cog of a fast rotating cogwheel or with the blades inside a turbomolecular pump. The disadvantage of disaggregated, micrometer-sized particles can be their unknown state of rotation and charge.

The thus disaggregated particles may have an initially velocity from the dispersion process which can be used. If not, they can be accelerated by gravity or suspended in an aerosol to be transported and then separated into gas and particles by a nozzle and/or a skimmer.

\subsection{Preparation and handling of ice particles}

In general icy samples have to be handeled and prepared in a different way than `warm' samples. Ice samples can be prepared from larger ice blocks by melting them to shape afterwards, they can be frozen inside molds or can be produced by freezing of water droplets in liquid nitrogen or a cold and dry gas environment. The former methods are often used for the production of larger scaled samples, whereas the latter can be used to create ice samples of a few millimeters and less. The production of large ice blocks without cracks is also challenging and can either be achieved by subsequently freezing multiple layers of water or by very slowly decreasing the temperatures.

Additionally, ice samples have to be handeled in a cold and dry environment e.g. cold nitrogen gas, cold rooms or cryogenic vacuum chambers, to avoid frosting of the surfaces that will alter the collisional properties. Whenever samples are loaded into experimental setups it has to be assured that the sample carriers are pre-cooled to the desired temperatures and that the tools are cold as well. Otherwise surface changes cannot be avoided.

\section{A review of impact experiments in the literature}\label{sec:review}

\begin{figure}[t]
    \begin{center}
        \includegraphics[width=\columnwidth]{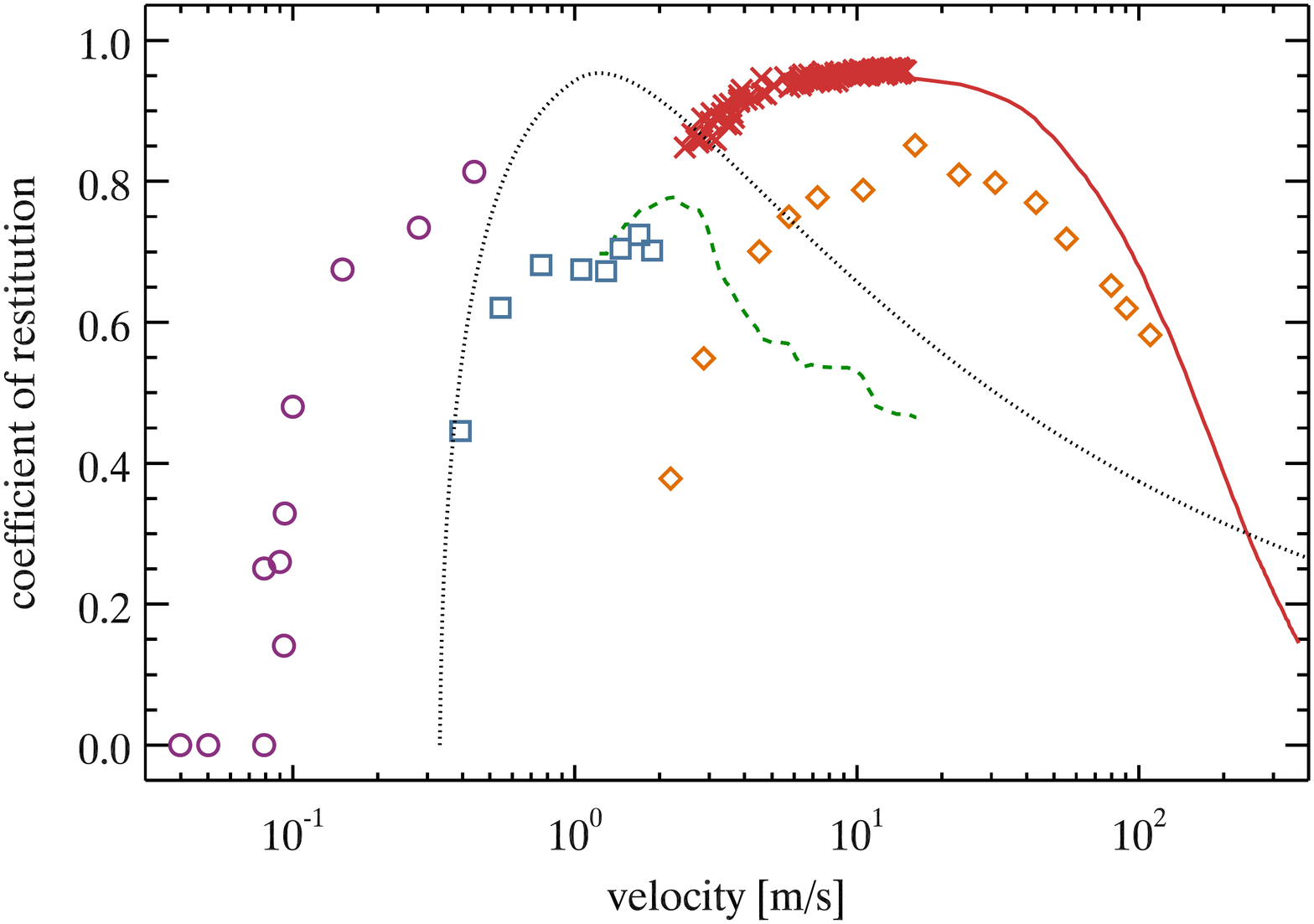}
        \caption{\label{fig:micrometer}Examples of micrometer particle collisions showing adhesive as well as plastic behavior. Red crosses and solid line \citet[latex, 1.3 \mum]{Dahneke:1975}, green dashed line \citet[silica, 1.2 \mum]{PoppeEtal:2000a}, orange diamonds \citet[fluorescein, 5 \mum]{WallEtal:1990}, blue squares \citet[steel, 10--65 \mum]{LiEtal:1999}, purple circles \citet[Ag-coated glass, 40 \mum]{KimDunn:2008}. The black dotted curve represents one arbitrary model curve of \citet{ThorntonNing:1998} for $v_\mathrm{y} = 3 \cdot v_\mathrm{s} = 1$ \ms.}
    \end{center}
\end{figure}

\begin{figure}[t]
    \begin{center}
        \includegraphics[width=\columnwidth]{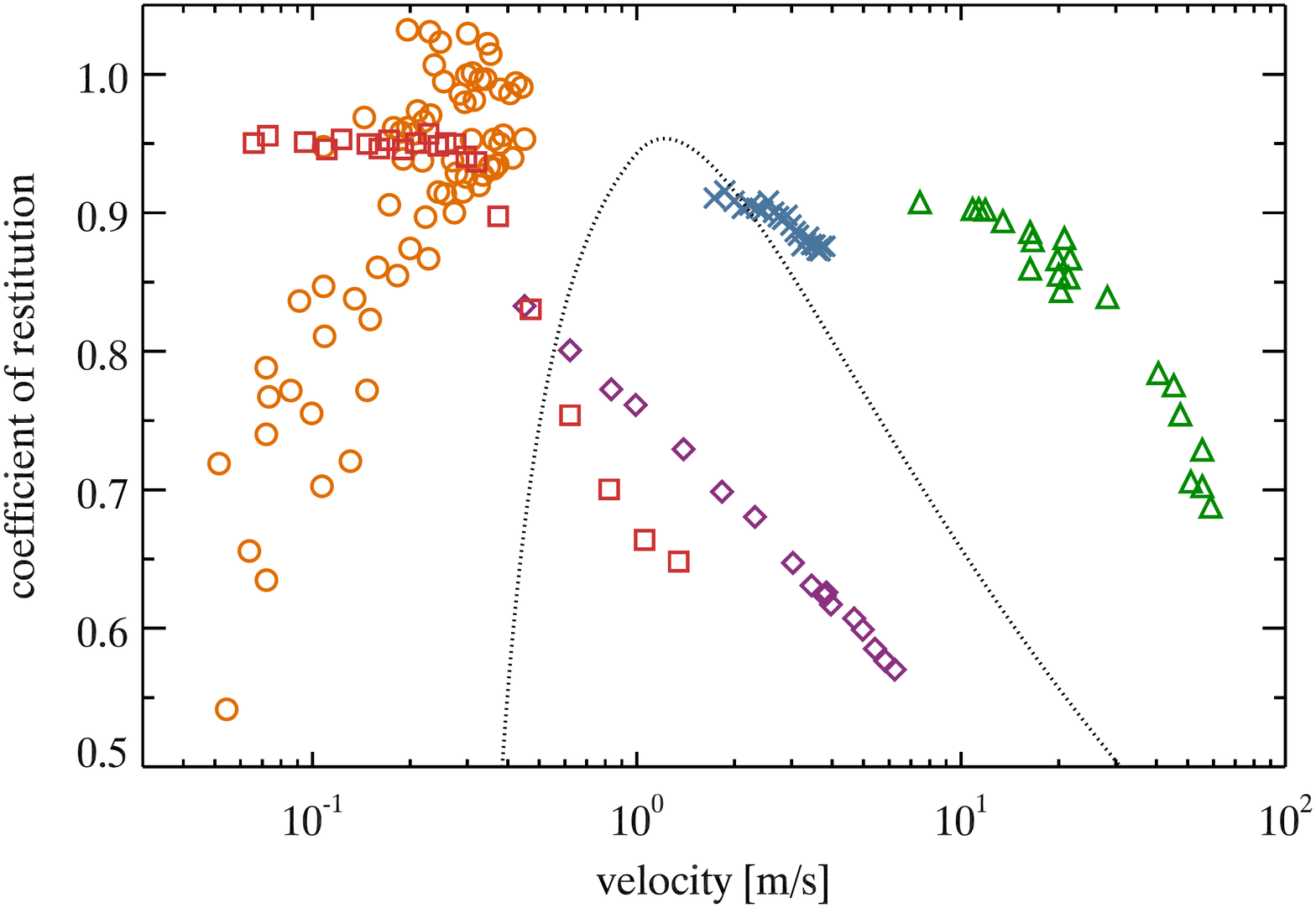}
        \caption{\label{fig:millimeter}Examples of millimeter particle collisions. Orange circles \citet[alumina, 5mm]{SoraceEtal:2009}, purple diamonds \citet[alumina, 5mm]{GorhamKharaz:2000}, red squares \citet[steel, 13 mm]{LifshitzKolsky:1964}, green triangles \citet[nylon, 7 mm]{LabousEtal:1997}, blue crosses \citet[steel, 5 mm]{SondergaardEtal:1990}. The black dotted curve represents one arbitrary model curve of \citet{ThorntonNing:1998} for $v_\mathrm{y} = 3 \cdot v_\mathrm{s} = 1$ \ms.}
    \end{center}
\end{figure}

In this section, we review the laboratory experiments on normal collisions which were available in the literature and present a velocity dependence of the coefficient of restitution. A list of all experiments is given in Table \ref{tab:overview_table}, which is roughly categorized into particles in the micrometer size range (i.e., below 1 mm), particles larger, and particles consisting of ice. To some level, this distinguishes also between experiments which show adhesive behavior (rising slope in Fig. \ref{fig:TN_plot}) and collisions where plastic behavior dominates (falling slope in Fig. \ref{fig:TN_plot}). Ice is treated separately because of the different techniques and problems with these experiments. The experiments with micrometer particles are described in Sect. \ref{subsec:micrometer}, those with millimeter particles in Sect. \ref{subsec:millimeter}, and the ice experiments are outlined in Sect. \ref{subsec:ice}. Examples of the coefficient of restitution data for the two size regimes are presented in Figs. \ref{fig:micrometer} and \ref{fig:millimeter}.

\begin{table*}[!t]
\small
\begin{center}
\caption{List of referenced experiments.}\label{tab:overview_table}
\begin{tabular}{cccccl}
\multicolumn{6}{c}{\textsc{Experiments with Micrometer Particles}}\\ \hline
regime & projectile material & target material & diameter [\mum] & velocity [\ms]  & reference\\ \hline\\[-0.9em]
\rowcolor[gray]{.9}S A P & polystyrene latex, & quartz & 1.3, 2.0 & 2 -- 375 & \citet{Dahneke:1973, Dahneke:1975}\\
\rowcolor[gray]{.9} & polyvinyl-toluene & & & &\\
A P & ammonium & molybdenum, PVF, & 2.6 -- 6.9 & 1 -- 100 & \citet{WallEtal:1990},\\
& fluorescein & silicon, mica & & & \citet{John:1995}\\	
\rowcolor[gray]{.9} A & Ag-coated glass & steel, copper, & 1 -- 30 & 2 -- 25 & \citet{DunnEtal:1995}\\
\rowcolor[gray]{.9} & & aluminum, PVF &&&\\	
A & steel & silicon & 10 -- 125 & 0.4 -- 4.6 & \citet{LiEtal:1999}\\
\rowcolor[gray]{.9} S A & Ag-coated glass & silica & 40 & 0.04 -- 0.44 & \citet{KimDunn:2008}\\
S P & silica & silica, silicon & 0.5 -- 1.2 & 0.2 -- 20 & \citet{PoppeEtal:2000a}\\
\\[-0.9em] \hline \\
\multicolumn{6}{c}{\textsc{Experiments with Millimeter Particles and Larger}}\\ \hline                     
                      regime   & projectile material                & target material      & diameter [mm]   & velocity [\ms]  & reference\\ \hline\\[-0.9em]
\rowcolor[gray]{.9}   C P      & steel                              & steel                & 12.7            & 6.5 -- 135      & \citet{LifshitzKolsky:1964}\\	
                      P        & steel, brass, glass, cork          & same                 & 15 -- 20        & 0.2 -- 5        & \citet{KuwabaraKono:1987}\\
\rowcolor[gray]{.9}   P        & steel                              & steel                & 1.29            & 5 -- 200        & \citet{KangurKleis:1988}\\
                      P        & steel                              & PMMA, aluminum       & 5, 10           & 1.7 -- 3.8      & \citet{SondergaardEtal:1990}\\
\rowcolor[gray]{.9}   C        & soda-lime glass,                   & either same          & 3, 6          & 0.3 -- 1.7      & \citet{FoersterEtal:1994}\\
\rowcolor[gray]{.9}            & acetate                            & or aluminum          &                 &                 & \\  
                      C        & polysterene latex,                 & same (free or glued) & 3 -- 5          & 0.2 -- 1.9      & \citet{LorenzEtal:1997}\\
                               & glass, acrylic, steel              & or aluminum          &                 &                 & \\  
\rowcolor[gray]{.9}   P        & nylon                              & same                 & 6 -- 25         & 0.7 -- 60       & \citet{LabousEtal:1997}\\
                      C        & tungsten carbite                   & piezo sensor         & 8               & 0.002 -- 0.023  & \citet{FalconEtal:1998}\\
\rowcolor[gray]{.9}   P        & alumina                            & aluminum alloy       & 5               & 0.5 -- 6.3      & \citet{GorhamKharaz:2000}\\
                      P        & alumina                            & polycarbonate        & 3.2             & 0.2 -- 2.2      & \citet{LougeAdams:2002}\\
\rowcolor[gray]{.9}   P        & brass, aluminum                    & same                 & 55              & 0.3 -- 1.3      & \citet{WeirTallon:2005}\\
P & steel & same & 25.4 & 0.4 -- 2.0 & \citet{StevensHrenya:2005} \\
\rowcolor[gray]{.9}                      P        & steel                              & sandstone            & 3               & 0.14 -- 2       & \citet{ImreEtal:2008}\\
A & iron & same & 4 & 0.01 -- 0.95 & \citet{GrasselliEtal:2009} \\
\rowcolor[gray]{.9}   A        & alumina, steel, acrylic            & same                 & 3, 4            & 0.05 -- 0.5     & \citet{SoraceEtal:2009}\\
                      C        & granite                            & same                 & 1000            & 0.1 -- 1.5      & \citet{DurdaEtal:2011a}\\ 
\rowcolor[gray]{.9} P & steel, brass, aluminum, delrin & steel, granite & 6 -- 19 & 0.001 -- 2 & \citet{KingEtal:2011} \\											
P & steel & glass & 6 & 0.2 -- 1.2 & \citet{MontaineEtal:2011} \\
\rowcolor[gray]{.9}   C        & soda-lime glass                    & same                 & 10              & 0.01 -- 1       & \citet{Beger_dipl}\\
\\[-0.9em] \hline \\ 
\multicolumn{6}{c}{\textsc{Experiments with ice particles}}\\ \hline
                      regime   & projectile material                & target material      & diameter [cm]   & velocity [\ms]  & reference\\ \hline\\[-0.9em]
\rowcolor[gray]{.9}   S P      & ice                                & same                 & 5 -- 50         & $10^{-5}-10^{-2}$& \citet{BridgesEtal:1984, BridgesEtal:1996},\\	
\rowcolor[gray]{.9}            &&&&& \citet{HatzesEtal:1988, HatzesEtal:1991}\\	
                      C        & ice                                & same                 & 2 -- 5          & 0.01            & \citet{DilleyCrawford:1996}\\
\rowcolor[gray]{.9}   C P      & ice                                & same                 & 0.28 -- 7.2     & 0.01 -- 10      & \citet{HigaEtal:1996, HigaEtal:1998}\\
                      C        & ice                                & same                 & 1.5, 1          & 0.06 -- 0.22    & \citet{HeisselmannEtal:2010}\\
\\[-0.9em] \hline
\end{tabular}\end{center}
Regime abbreviations denote: A: adhesive regime; P: plastic regime; S: sticking velocity; C: regime with constant coefficient of restitution
\end{table*}

\subsection{Experiments with micrometer particles}\label{subsec:micrometer}

\citet{Dahneke:1975} presented direct time-of-flight measurements of 1.27 \mum\ polystyrene latex and 2.02 \mum\ polyvinyl-toluene microspheres in a collimated beam striking a polished quartz target. The incident particles crossed two light barriers with a known distance, struck the target and were rebound to cross the light barriers again. The velocity was controlled by the residual air pressure in the last of a series of vacuum chambers which decelerated the microspheres; possibly charged particles were precipitated by an electric field before entering this last vacuum chamber. It is noteworthy that the rebounding velocity vector was found to be normal to the target within a diffusion angle as small as 1° \citep[in contrast to to][below]{PoppeEtal:2000a}. With these measurements, they provided low-velocity (2 -- 15 \ms) data on the coefficient of restitution in the adhesive regime as well as high-velocity data (15 -- 375 \ms) in the plastic regime for the 1.27 \mum\ latex spheres and an intermediate regime for the 2.02 \mum\ polyvinyl-toluene particles. The curve for the plastic regime shows no scatter and it is not exactly described how this was achieved. The data was shown however in \citet{Dahneke:1995} and are very smooth. With the experimental setup described in \citet{DahnekeFriedlander:1970}, also using a collimated particle beam of the latex particles, \citet{Dahneke:1973} provided measurements on the sticking velocity of these. To achieve this, he measured the distribution of particles in the beam and the fraction of particles sticking to the target to indirectly deduce the sticking probability function. The critical sticking velocities thus obtained were 0.83 and 0.92 \ms\ for for stainless steel and polished quartz targets, respectively. The sticking velocities were higher than expected and one possible explanation is the coating of the latex particles with a layer of a dried stabilizing agent with an expectedly different surface energy.

\citet{WallEtal:1990} presented a sophisticated setup of freshly producing ammonium fluorescein particles, transporting them to a 'bounce cell' with different target materials and measuring incoming and rebound velocities with a dual-laser-beam Doppler system. The fluorescein particles (3 -- 7 \mum\ diameter) produced in a vibrating-orifice aerosol generator were analyzed to have surface roughnesses of 3 nm and targets were also chosen with greatest care considering surface conditions: molybdenum was hand-polished with 1200-mesh silicon carbide particles; silicon was cut from an electronics-grade, single-crystal wafer; muscovite mica was handled with care to maintain the crystal layer; polyvinylfluoride (PVF, Tedlar) was chosen as a soft, deformable target. Surfaces were examined by scanning electron microscopy (SEM) and laser interferometer and roughnesses were as good as 5 nm for molybdenum on the microscale (still with 0.3 \mum\ asperities) and better than 0.1 \mum\ (resolution limit of their SEM) for silicon and mica targets. Velocities were taken by Doppler shift measurements 10 \mum\ above the target to reduce the influence of deceleration from the residual gas and not processed for single impacts but mean values of 250 individual collisions were taken instead. As the impacts were upside down, bouncing secondary collisions with velocities as small as 2 \ms\ could be included. This yielded a measurement range of coefficients of restitution from 2 to 115 \ms, resolving the adhesive as well as the plastic branch of the $e(v_\mathrm{i})$ curve. Additional data of this work, especially those on different particle sizes,  were later presented in the review article by \citet{John:1995}.

The experimental setup of \citet{DunnEtal:1995} had some similarities to the one of \citeauthor{WallEtal:1990} in that velocities were also measured with a similar Doppler system right before the target plate. In their experiments, the particles were glass microspheres (1 -- 30 \mum) coated with a 500 \AA\ thick silver layer, which were dispersed in an electrostatic particle disperser \citep[described by][]{OlansenEtal:1989} with subsequent discharging. Also steel and nickel particles were mentioned but no quantitative results were presented. After dispersing, the spheres had exit velocities of up to 20 \ms\ and were further vertically accelerated to achieved velocities from 2 to 25 \ms. Targets were stainless steel, copper, and aluminum, which were hand polished to achieve smooth, mirror like finishes verified with an SEM (steel as smooth as 40 \AA, others with 1 \mum\ irregularities). Some steel targets were coated with a 2 \mum\ layer of siloxane, a fifth surface type, PVF, was handled as a thin film that was glued on an aluminum mount. It is noteworthy that all particles had a conducting surface to avoid spot charges and the charge was explicitly measured on the grounded, also conducting target. The authors concluded that charge effects did not have an influence on the results on the coefficient of restitution. Moreover, in contrast to \citeauthor{WallEtal:1990}, \citeauthor{DunnEtal:1995} were able to correlate the incident and rebound of \textit{individual} particles to measure the coefficient of restitution of individual collisions instead of mean values. Mean values were taken only after that. In their data, a clear effect of adhesive energy dissipation for velocities below approx. 10 \ms\ can be seen for each target. Moreover, the level of relatively constant coefficient of restitution was the lowest for the softest material (PFV) and the highest for the hardest material (steel).

\citet[from the same group]{LiEtal:1999} used the same experimental setup as \citeauthor{DunnEtal:1995} with the major difference that the particles were accelerated only due to gravity and the drop height yielded velocities in the range from 0.4 to 4.6 \ms, i.e., slower than those of \citeauthor{DunnEtal:1995} It is therefore not clear whether they used the same particle disperser, although this is well possible as the particles were different. Those were stainless steel microspheres of two kinds with size distributions between 10 -- 65 \mum\ and 60 -- 125 \mum\ and smoothness as good as the resolution of their SEM. For the target they used a plane silicon crystal with surface-asperity heights within 10 \AA\ standard deviation. Although the velocities were significantly smaller than those of \citeauthor{DunnEtal:1995}, coefficients of restitution in the adhesive regime are still as high as 0.4 due to the larger particle size.

A third publication from this group, the article of \citet{KimDunn:2008}, provided direct imaging results of micro particle collisions. 40 \mum\ glass microspheres coated with Ag were dispersed and vertically dropped on a silica target plate. The collisions under atmospheric conditions were resolved with a high-speed camera in back-light illumination of which an example is presented in the paper. A small number of 11 collisions at velocities from 0.04 to 0.44 \ms\ clearly shows the adhesive branch of the $e(v_\mathrm{i})$ curve with three sticking instances at the low velocities. Particle and target surface conditions were not further described so it must be assumed that at least the Ag-coated projectile particles were similar to those having been used by \citeauthor{DunnEtal:1995}

Experiments with 0.5 and 1.2 \mum\ diameter silica spheres were performed by \citet{PoppeEtal:2000a}. The particles were deagglomerated with a fast rotating cogwheel (ca. 50 \ms\ circumference velocity) and then brought to collision with a flat silica or silicon-wafer target at a velocity of 0.2 to 20 \ms. Trajectories of impacting and rebounding particles were observed by a high-speed camera in the forward scattered light of a pulsed laser. Due to this optical setup, a strip of approx. 10 to 40 \mum\ next to the target could not be observed. The impacting particles possessed a charge of few elementary charges, which was studied in a second paper by \citet{PoppeEtal:2000b} and were found to significantly charge up due to collisional tribocharging with the target. In contrast to the findings of \citet{Dahneke:1975} the rebound angles exhibit a significant scatter, possibly due to an unknown rotation around their own axis. Also the coefficient of restitution shows a large scatter, but the averaged data shows a clear decrease with increasing velocity. The authors present detailed measurements on the sticking velocities for three of the four projectile-target combinations. The adhesive branch of the $e(v_\mathrm{i})$ curve (cf. Fig. \ref{fig:TN_plot}) can unfortunately not be concluded from the presented data. Surface conditions from measurements with an atomic force microscope (AFM) were reported in their Table 1 for targets and projectiles and were in a range of 0.15 to 1 nm in root mean square deviation from the average surface for a scanned area of $50 \times 50\ \mathrm{nm}^2$.

\subsection{Experiments with millimeter particles}\label{subsec:millimeter}

\citet{LifshitzKolsky:1964} performed experiments with 12.7 mm steel balls impacting on a 5 cm thick mild-steel block. The steel ball was suspended on a thin long wire on which it was pulled aside to horizontally collide with the steel block in a series of rebound collisions. The friction of this pendulum (1 -- 2 \% energy loss per swing) was measured in free oscillation without a steel block and corrected from the results. Camera observations of the bouncing steel ball yielded the coefficient of restitution for a range of velocities from 0.07 to 1.4 \ms\ which showed to be constant for velocities less than 0.3 \ms\ where the material started to yield and show a decline of the coefficient of restitution for higher velocities. The authors also discussed the effect of surface roughness: for their initial conditions, the coefficient of restitution in the low-velocity (elastic) range was below 0.82 with a significant scatter. In two polishing steps, they were able to reduce the scatter and increase the constant coefficient of restitution to 0.87 and 0.95. The final surface conditions are described as highly reflecting mirror finish, however with small pits observable under an optical microscope.

The sparsely described experiments of \citet{KuwabaraKono:1987} were performed on a 2 m long pendulum between steel, brass, glass, and cork spheres of 15 to 20 mm diameter. Additionally, they used cork spheres with a lead core and a cork plate fixed to a heavy iron block. Velocities were obtained using a video camera and are in a range from 0.2 to 5 \ms. Some of the datasets show a decrease of the coefficient of restitution with increasing incident velocity.

The article of \citet[in Russian]{KangurKleis:1988} was not accessible to us but is featured by \citet{KimDunn:2007}. The experiments of a 1.29 mm diameter stainless steel ball impacting a massive stainless steel plate were performed at a velocity from 5 to 200 \ms. The coefficient of restitution is falling from 0.7 to 0.1 over that range.

\citet{SondergaardEtal:1990} presented vertical impacts of steel and bronze ball bearings onto plates of polymethyl methacrylate (PMMA, lucite) or aluminum. One focus of this work was on the effect of the plate thickness but one series of experiments with steel ball bearings of 5 and 10 mm diameter colliding with a 1.27 cm thick PMMA plate shows the weak velocity dependence of the coefficient of restitution for these collisions in a range from 1.7 to 3.8 \ms. The only information about the surface conditions was that the steel balls were precision ball bearings.

An experimental setup to study free particle-particle collisions without rotation is presented by \citet{FoersterEtal:1994}. Soda-lime glass spheres of 3 mm diameter or cellulose acetate spheres of 6 mm diameter are vertically aligned and dropped with a slightly different timing. The top sphere is released first and the velocity is determined by the time difference of releasing the second particle. Beside the mutual collisions, impacts on 64 mm thick aluminum plates were studied with the same setup. With stroboscopic images of the particles before and after the collision, the authors determined the incident and rebound velocities of the oblique impacts. A model of \citet{Walton:1994} and the balance between linear and angular moments determines the normal and tangential coefficient of restitution, which are given as a constant for the studied range from 0.3 to 1.7 \ms\ in normal velocities $v_\mathrm{n}$ -- the absolute velocity $v_\mathrm{i}$ was constant for either sphere-sphere or sphere-plate collisions and only the incident angle or impact parameter was changed. The raw data of $e(v_\mathrm{n})$ is not presented and the constant value over the considered velocity range is a model assumption. Surface conditions are stated as polished and grade 200 for the soda-lime glass, ashed for the acetate and machine finish for the aluminum plate.

The same setup is used by \citet{LorenzEtal:1997} using lead-free glass, polystyrene, acrylic, and steel spheres of 3, 4, 4, and 5 mm diameter, respectively. Some experiments are reported in which the bottom spheres were fixed on a stationary aluminum plate. The surface conditions are such that they represent affordable, normal quality spheres. An indication for the quality is the deviation from sphericity of 1.1 \% of the diameter for the glass beads. For another application focused study, they used 'spent' glass beads, which were circulated in a chute facility for 20 hours and experienced considerable abrasion. The difference can be seen in the scatter of the data, but again the raw $e(v_\mathrm{n})$ data is not presented but only the mean valued deduced from the model. The velocity range in $v_\mathrm{n}$ between 0.2 and 1.9 is similar to the one of \citeauthor{FoersterEtal:1994} Besides the free collisions and the collisions with a plate, also collisions with a sphere glued to the bottom plate were studied.

Experiments from the same group are described in \citet{LougeAdams:2002}. The apparatus is similar to the one described by \citeauthor{FoersterEtal:1994} and the 3.2 mm alumina beads collide with a polycarbonate plate at oblique impact angles. Again based on the model of \citeauthor{Walton:1994}, the authors concluded the normal coefficient of restitution from the oblique impact and presented a velocity dependence of the coefficient of restitution. Two datasets of different drop heights yield two different absolute velocities and with a variation of the impact angle a range of normal velocities. However, these two datasets do not satisfactorily collapse and the authors conclude that the change in the coefficient of restitution cannot solely be attributed to the variation of the normal velocity. They partly explain this by an expectedly asymmetric deformation of the sphere during the contact. It should be noted that coefficients of restitution of normal impacts and the normal components from oblique impacts were highly comparable in the results presented by \citet[see below]{GorhamKharaz:2000}. However, due to the geometry the data on the coefficient of restitution of \citeauthor{LougeAdams:2002} lie mainly above unity (0.9 -- 1.3), in contrast to \citeauthor{GorhamKharaz:2000}. The alumina spheres were obtained from Hoover Precision who provided either grade 10 or 25 surface quality at the time of writing this article.

\citet{SoraceEtal:2009} developed a precise pendulum setup to study the collision of small but macroscopic spheres of alumina, steel (both 3 mm diameter) and acrylic (4 mm) at velocities between 0.05 and 0.5 \ms. The spheres were each suspended on two spectra fibers with precisely tuned length and released from a certain angle to collide several times. According to the authors, the coefficient of restitution did not change after several collisions because the spheres were to small to be sufficiently flattened in the collisions (i.e., no wear; in contrast to experiments with heavier particles by \citeauthor{FalconEtal:1998} or \citet[see below]{WeirTallon:2005}). With a photogate they measured the time between two successive collisions to calculate the highest oscillation point and take a picture. The resultant coefficients of restitution for the higher velocities were consistent with measurements of \citeauthor{LorenzEtal:1997} and showed a typical adhesive decline for velocities below approx 0.25 \ms. Fitting their results with models of \citet{ThorntonNing:1998} and \citet{BrilliantovEtal:2007}, the extrapolated sticking velocities were found to be much higher than expected.

Free particle-particle impact experiments at higher velocities were performed by \citet{LabousEtal:1997} using 6.35, 12.7, and 25.4 mm diameter nylon spheres and a video imaging system. The particles were fit into a tube, held by an underpressure and then pushed out by an instantly applied overpressure to achieve relative velocities between 0.7 and 60 \ms. For velocities higher than 10 \ms\ the coefficient of restitution declined from a relatively high and constant level, which indicates plastic behavior. Unfortunately, the authors do not provide detailed information on the surface conditions of their nylon spheres.

\citet{FalconEtal:1998} used a similar particle release mechanism as \citeauthor{FoersterEtal:1994} (holding the particle with an underpressure) and studied the collision between 8 mm diameter, non-rotating tungsten carbite beads and a PCB 200B02 piezoelectric sensor mounted on a 2 cm thick duraluminum (an aluminum alloy) plate. The collisions were normal and vertical, i.e., downward incident velocity. They studied multiple bounces and recorded the data from the piezoelectric sensor with a digital oscilloscope with a time resolution of 10 ns which only recorded the last and thus slowest collisions. To achieve a sufficient number of bouncing collisions, the target plate was carefully adjusted in the horizontal direction and the spheres had to be precise enough. The tungsten carbide beads described in the paper had a tolerance of $\pm 1$ \mum\ and a sphericity of $\pm 0.8$ \mum, surface conditions were not described. In the studied velocity range from 0.02 to 0.23 \ms\ they observed a constant coefficient of restitution around 0.97 for the higher velocities and an increasing scatter whilst a slightly decreasing trend for velocities below 0.1 \ms. Although this could be expected to be indicative for adhesion, the authors argue in detail that this effect can be attributed to gravity.

In the experiments described by \citet{GorhamKharaz:2000}, a 5 mm diameter alumina bead is dropped onto an aluminum alloy plate from drop heights between 1 cm and 2 m (0.45 to 6.3 \ms). The beads are precisely released in a similar fashion as \citeauthor{FoersterEtal:1994} and cross a light barrier to trigger a camera and a stroboscopic light for achieving very reproducible measurements. This requires very precise surface conditions of projectile and target, which can be concluded from the data but is not described (only the deviation from sphericity of the alumina sphere is stated as $< 0.12$ \mum). The focus of the paper is on the angular dependence of the coefficient of restitution and the validation of the model of \citet{MawEtal:1976} but also the velocity dependent coefficient of restitution for normal collisions exhibits a low scatter and shows a clearly plastic behavior. The coefficient of restitution from the normal velocity component of a series of oblique impacts is indistinguishable from those of a series of normal impacts.

The rather theoretically focused work of \citet{WeirTallon:2005} features some collision experiments to verify their theory. These are experiments with a pendulum using a pair of 3.8 m long thin wires for each particle. The two colliding particles were either 55 mm diameter brass and aluminum beads of one of these beads and 50 mm diameter, 45 mm length cylinder colliding with the flat plane. One sphere was raised and released and the incident and rebound velocities were calculated from the release height and the furthest extend of travel after the collision, recorded by a video camera. In the velocity range from 0.3 to 1.3 \ms, a decreasing trend for the coefficient of restitution can be observed, while it is not clear whether the data presented are mean values or single impacts. The surface conditions of the spheres were not described.

In the experiments of \citet{StevensHrenya:2005} a pendulum was used to study collisions between either stainless steel or chrome steel balls of 25.3 mm diameter. The velocities were measured with a pair of light barriers for each particle and they also measured the contact time of the electrically connected spheres via resistivity measurement. While  \citeauthor{StevensHrenya:2005} provide the exact names for their materials (steel grade 316 and AISI 52100, respectively) and, with this, the mechanical properties, they do not make a statement about the surface conditions.

\citet{ImreEtal:2008} dropped 3 mm ball bearing steel balls onto natural polished sandstone targets (see \citeauthor{ImreEtal:2008} for a detailed description of the target material). The steel projectile was released by switching off either an electromagnet or a vacuum (both methods were mentioned in the paper and it is not clear which was used for this series). The drop heights of 1 to 200 mm determine an impact velocity of 0.14 to 2 \ms\ and although multiple bounces were taken into account, velocity dependent coefficients of restitution are only given for very few discrete velocities. The coefficients of restitution were determined from the flight times between two bounces, recorded with an acceleration sensor glued to the target material. Different target materials were also studied but these were performed only for a single constant velocity where the coefficient of restitution was expected to be at a constant plateau before the onset of plastic dissipation.

\citet{GrasselliEtal:2009} performed an experiment under the reduced gravity condition of a parabolic flight. They studied an ensemble of 4 mm diameter iron beads contained in a glass cell. This cell was vibrated to distribute the particles in the cell and the vibration was then turned off after a few seconds. The particles collided and due to this energy dissipation became gradually slower, allowing for collision velocities down to 1 \mms. For slow velocities below 20 \cms\ the coefficient of restitution is decreasing. A general drawback of this kind of experiment is however that the rotation of the particles, which contributes to the coefficient of restitution, is unknown. The surface conditions of the iron beads were not specified.

Impact experiments with the largest samples we are aware of were performed by \citet{DurdaEtal:2011a}. Suspended on approx. 15 m long cables on two cranes were two granite spheres of 1 m diameter. These were milled from massive blocks and the surfaces were described as relatively smooth but not being polished. From the feasible angles to raise the 1.3 ton particles, \citeauthor{DurdaEtal:2011a} achieved a velocity range from 0.1 to 1.5 \ms\ and found a relatively constant coefficient of restitution of $0.8 \pm 0.06$, where the standard deviation is also more or less comparable to the measurement accuracy.

With a vacuum release mechanism, \citet{KingEtal:2011} dropped steel, brass, aluminum, and Delrin (POM-H) spheres of diameters from 6.24 to 18.76 mm (most were 9.4 mm) onto solid plates. This plate was either a steel disk of 22 cm diameter and 2.5 cm thickness or a heavy optical table from granite ($120\times80\times30\ \mathrm{cm^3}$). In most experiments the contact with the plate was detected with an accelerometer, which was attached 5 cm from the impact point. For the lowest velocities down to 1 \mms\ they used the steel target and conductive balls, which were electrically connected (a detailed description on this is not given). The coefficient of restitution in both cases follows from the time differences between the individual collisions. For the brass balls (which were most used in that study) a surface topography is presented in their Fig. 3. The mean deviation is better than 0.5 \mum. They moreover describe how they frequently change the spheres and surface grind the steel plate. As a result, \citeauthor{KingEtal:2011} find a coefficient of restitution which they describe as non-monotonic. While plastic  dissipation is likely for velocities larger than 0.1 \ms, the behavior at lower velocities cannot easily be understood.

A very large set of data, including $2\cdot10^6$ collisions is presented by \citet{MontaineEtal:2011}. Using a vacuum release mechanism, they studied the collision of a 6 mm diameter stainless steel ball bearing with a glass target ($30\times10\times1.9\ \mathrm{cm^3}$). Again, a piezo sensor connected to the glass plate measured the time difference between successive collisions and thus collision velocities and  coefficients of restitution of 90 -- 100 successive collisions. After such a set of data, the sphere was picked up by a robot arm and dropped from the same height but at a random position within the target plate's center. Experiment were conducted at air pressure but it is described that temperature and humidity were kept constant. The surface conditions is described as having tiny scratches. Moreover, since the same sphere was used for all collisions, sphere and plate were analyzed with a Scanning Electron Microscope at the beginning and the end of these experiments without revealing a significant difference. an interesting result is the observed scatter of the coefficient of restitution. For a constant velocity, the mean value is very significant but the data smoothly scatter around this value.

\citet{Beger_dipl} performed free collision experiments with a two-particle release mechanism inside a 1.5 m long vacuum glass tube. Two high-speed cameras were synchronously falling outside the glass tube to observe the collision; details of the setup are provided by \citet{BeitzEtal:2011}. The long free fall length and the falling camera allow a collision velocity from 0.01 \ms\ compared to the 0.2 \ms\ of \citeauthor{FoersterEtal:1994} without using a glass tube. The highest velocity that was studied by \citeauthor{Beger_dipl} was 1 \ms. The normalized impact parameters were mostly around 0.1 and always smaller than 0.25 and they used 10 mm soda-lime glass beads which were either smooth or etched. The surfaces were scanned with a laser displacement meter and show a roughness of $\pm 0.1$ \mum\ for the smooth spheres and $\pm 5$ \mum\ for the etched ones. Additionally, the smooth spheres show about one asperity of 1 \mum\ height per measured profile line of 100 \mum\ length. The data show coefficients of restitution of 1.01 for the smooth spheres at 1 \ms, with a scatter of less than 0.01, which can only be explained by a systematic error (e.g., due to a slight rotation before the collision). However, there is a statistically significant trend for decreasing coefficient of restitution with decreasing velocity (0.93 at 0.01 \ms), which does not really compare to the adhesive part of the red curve in Fig. \ref{fig:TN_plot}. The etched spheres generally have a smaller coefficient of restitution, which is evident in the mean values, but show the same trend.

\subsection{Experiments with ice particles}\label{subsec:ice}

\citet{BridgesEtal:1984, BridgesEtal:1996} and \citet{HatzesEtal:1988, HatzesEtal:1991} present results obtained from experiments colliding water ice spheres with a flat brick-shaped ice wall by means of a sophisticated compound disk pendulum. The collision velocities and the respective coefficients of restitution were obtained from frequency measurements of the pendulum using a capacitive deplacement detection. The used samples were of radii of curvature 2.5, 5, 10, and 20 cm and were collided inside a cryostat apparatus at velocities between $10^{-5}$ and $10^{-2}$ \ms, temperatures ranging from 90 to 170 K. The sample surfaces were categorized into three different surface conditions, namely (1) smooth and frost free, (2) roughened through sublimation, and (3) frost covered, where the latter was quantified to comprise a 10 -- 30 \mum\ thick frost layer. The authors reported a decay of the coefficient of restitution with increasing collision velocity and derived exponential and power-law descriptions from the fits to their data. They also found that surface roughness, e.g., frost layers, significantly reduced the coefficient of restitution and quantified this effect to be as large as 30 \%. Frost was also considered to be responsible for the scatter of their data. The influence of the frost layers showed to decrease with the number of collision contacts at the same spot indicating a local compaction. 

\citet{DilleyCrawford:1996} used a modified deep freezer that comprised the bottom end of a bifilarly suspended string pendulum to collide ice spheres of 2, 2.4, 3, and 5 cm diameter with a solid ice block of 4 kg mass. The pendulum of 10 m length allowed for central collisions at 253 K temperature. The collision velocities were obtained by measuring the pendulum amplitude from a video recording in slow motion playback  by means of a ruler. Prior to the collision experiments the sample spheres were collided with the target block for a couple of times to compress frost layers that have accumulated  on the surface of the samples due to humidity of the ambient air. A detailed description of the surface properties was not given. The authors report data that was obtained for collision velocities of  approx. 0.01 \ms\ and they observed a  mass-dependent decrease of the coefficient of restitution with decreasing mass. The results were derived from mean values, because the varying humidity and ambient temperatures during the experiments caused significant scatter of the obtained data as a result of varying frost conditions. A velocity dependency was modeled using a model by \citet{Dilley:1993} that was derived to fit the data of \citet{HatzesEtal:1988}. \citeauthor{DilleyCrawford:1996} compared the modeled behavior of $e$ with control experiments using steel spheres and a steel brick. 

\citet{HigaEtal:1996, HigaEtal:1998} present results from drop bar experiments of polycristalline ice spheres. They used a bar to push the sample out of a cooling mold and thus initiated the impacts of smooth and frost-free ice spheres (0.28 -- 7.2 cm diameter) on polished ice blocks ($10\times 10\times 10\ \mathrm{cm^3}$, $28\times 28\times 28\ \mathrm{cm^3}$ and $9\times 9\times 3.5\ \mathrm{cm^3}$), respectively. The collision velocities ranged from 0.01 to 10 \ms\ and were observed at temperatures between 113 and 215 K using a cryostat apparatus or  245 to 269 K when being performed in a cold room. All measurements of the coefficients of restitution were obtained either by high-speed camera recording (200 or 500 fps) resolving the collisions or by acoustic emission sensors providing the time differences between consecutive bounces. The authors divide their data into three categories: (1) non-cracked spheres, (2) cracked spheres, and (3) fragmenting collisions. They found that the critical fracture velocity increases with decreasing temperature and is constant below 215 K. Below a critical velocity $v_\mathrm{c}$ the coefficients of restitution showed no velocity dependence, whereas it decreases according to $e = e_\mathrm{qe} ( v/v_\mathrm{c} )^{-\log(v/v_\mathrm{c})}$ with increasing velocity above $v_\mathrm{c}$. The factor $e_\mathrm{qe}$ describes the coefficient of restitution of the quasi-elasitic regime. A temperature dependence could not be observed, but a size dependence of the coefficient of restitution was observed in the quasi-elastic regime below $v_\mathrm{c}$. \citet{HigaEtal:1996} also conducted experiments with frosted spheres clearly showing a decreases of the coefficients of restitution with decreasing collision velocity, which contradicts the results of \citet{HatzesEtal:1988}.

\citet{HeisselmannEtal:2010} studied free individual collisions of water ice spheres under microgravity conditions at collision velocities of 0.06 to 0.22 \ms\, where the normalized impact parameters varied from 0 to 0.5. The particles of 1.5 cm diameter showed a mean coefficient of restitution of 0.45 with uniformly distributed values ranging from 0 to 0.84. The data shows large scatter that is explained by slight variations of thin frost layers on the particles' surfaces. A comparison to glass spheres with a rough etched surfaces (same as \citeauthor{Beger_dipl}) shows a similar behavior with comparable scatter of the data.

\section{Discussion} \label{sec:discussion}

In this section, we will draw a brief comparison between the described experiments and the general shape of the coefficient of restitution curve provided by \citet{ThorntonNing:1998}. This is just intended to put the data into a context while the full analysis can then be found in \citet{KrijtEtal:preprint_a, KrijtEtal:preprint_b}. We saw that the surface conditions of the samples can play a decisive role for the value as well as for the scatter of the coefficient of restitution. We will therefore provide further information (and speculation if necessary) on the surface conditions of the listed experiments.

\subsection{Qualitative consistency between experimental data and theoretical model} \label{subsec:discussion_comparison}

In Figs. \ref{fig:micrometer} and \ref{fig:millimeter}, we plotted a model curve of \citet{ThorntonNing:1998} as the black dotted line for a qualitative comparison with the provided data. The parameters are chosen as $v_\mathrm{y} = 3 \cdot v_\mathrm{s} = 1$ \ms, only to fit into the chosen plot range. As mentioned earlier (Sect. \ref{sec:introduction}), the curve can be shifted in the horizontal direction without changing the overall shape. This only involves the yield velocity $v_\mathrm{y}$, which is a material constant for any given sphere. Any change in the sticking velocity $v_\mathrm{s}$ will alter the shape of the maximum as illustrated in Fig. \ref{fig:TN_plot}. Knowing this, we can use this curve to compare it to the experimental data in Figs. \ref{fig:micrometer} and \ref{fig:millimeter}.

For the data with micrometer particles (Sect. \ref{subsec:micrometer} and Fig. \ref{fig:micrometer}) we first see that the rise of the coefficient of restitution around the sticking velocity is steep and rather well represented by the model curve \citep[especially evident for the data of][purple circles]{KimDunn:2008}. The maximums of the data of \citet[red crosses and solid line]{Dahneke:1975} as well as \citet[orange diamonds]{WallEtal:1990} look rather broad compared to the model. Even in the case of \citeauthor{Dahneke:1975}, the maximum is well below unity and not well represented by any curve with $v_\mathrm{y} \gg v_\mathrm{s}$ (compare red curve in Fig. \ref{fig:TN_plot}). The data for mean coefficients of restitution from \citet[green dashed line]{PoppeEtal:2000a} nicely fits the slope in the plastic regime and it even looks like there is a significant maximum. However, although this maximum is close to the sticking velocity of 1.1 -- 1.3 \ms\ measured by \citeauthor{PoppeEtal:2000a} it should be noted that the standard deviation of the data around the maximum is around 0.1 and thus still in the same order as the curving itself. So this might be physical while not statistically significant.

For the collisions of millimeter-sized particles or larger (Sect. \ref{subsec:millimeter} and Fig. \ref{fig:millimeter}), the slopes on the plastic side mostly fit to the model curve. The rather flat appearance of the data of \citet[blue crosses]{SondergaardEtal:1990} stretching a rather narrow range in velocity might be explained if it is close to the maximum, i.e., close to the yield velocity. Otherwise, there is no data showing a sharp maximum and the only plateau that can be seen is the one of \citet[red squares]{LifshitzKolsky:1964}. This would not fit well to the model curve as the coefficient of restitution is around 0.95. \citeauthor{LifshitzKolsky:1964} expected that the coefficient of restitution could have been increased when further polishing the surface -- they saw an increase in the coefficient of restitution when improving the surface quality. If this is true, a perfectly smooth surface should have a coefficient of restitution very close to unity and could be described with a model curve with $v_\mathrm{y} \gg v_\mathrm{s}$ and for velocities less than the yield velocity. The yield velocity should clearly be associated with the kink in the data of \citeauthor{LifshitzKolsky:1964}. There are some other data not featured in Fig. \ref{fig:millimeter} which also show a constant coefficient of restitution over a certain velocity range \citep[e.g.,][and others]{FoersterEtal:1994, FalconEtal:1998} below unity which are not explained by this model.

The increasing coefficients of restitution with increasing velocity in the data of \citet[orange circles]{SoraceEtal:2009} could in principle well be described by a curve with $v_\mathrm{s} \sim 0.06$ \ms, as it has been shown by these authors. However, this is rather surprising as this would require a surface energy of 100 $\mathrm{J\,m^{-2}}$  for alumina, while the expected surface energies are one to two orders of magnitude smaller. It is not expected for two 3 mm alumina spheres to stick to each other at velocities around 0.05 \ms. This result was also surprising to \citeauthor{SoraceEtal:2009} and still lacks a good explanation (the same conclusion holds for the steel and acrylic spheres of \citeauthor{SoraceEtal:2009}).

This section was intended give a brief overview on the appearance of a selection of experimental data compared to the model of \citet{ThorntonNing:1998}. A deeper and quantitative analysis and a comparison to other models will be given by \citet{KrijtEtal:preprint_b}. Additionally, fits to these data will yield material parameters, which can then be compared to the expected values as discussed here in the example of \citeauthor{SoraceEtal:2009}

\subsection{Surface Conditions}\label{subsec:discussion_surface}

\begin{table*}[t!]
\vspace*{0.35cm}
\small
\begin{center}
\caption{Surface conditions of the referenced experiments in the same chronological order as in Table \ref{tab:overview_table}.}\label{tab:surface_table}
\begin{tabular}{ll}
  \multicolumn{2}{c}{\textsc{Experiments with Micrometer Particles}}\\ \hline
                    reference                                   & surface condition\\ \hline\\[-0.9em]
\rowcolor[gray]{.9} \citet{Dahneke:1973, Dahneke:1975}          & target 'polished', very normal rebound (indication for good surface)\\
                    \citet{WallEtal:1990}, \citet{John:1995}    & projectiles smooth down to 3 nm, targets featureless at 0.1 \mum\ resolution\\
                                                                & (silicon, mica) or 5 nm with 0.3 \mum\ asperities (molybdenum)\\
\rowcolor[gray]{.9} \citet{DunnEtal:1995}                       & steel targets 'virtually flawless' at 40 \AA, others 1 \mum\ irregularities\\
                    \citet{LiEtal:1999}                         & projectiles 'smooth' within SEM resolution$^{(1)}$, target 10 \AA\ standard deviation\\
\rowcolor[gray]{.9} \citet{KimDunn:2008}                        & not described\\
                    \citet{PoppeEtal:2000a}                     & AFM measurement, 0.15 -- 1 nm root mean square\\
\\[-0.9em] \hline \\
\multicolumn{2}{c}{\textsc{Experiments with Millimeter Particles and Larger}}\\ \hline                     
                    reference                                   & surface condition\\ \hline\\[-0.9em]
\rowcolor[gray]{.9} \citet{LifshitzKolsky:1964}                 & 'highly reflecting mirror finish' with 'small pits' under optical microscope$^{(2)}$\\	
                    \citet{KuwabaraKono:1987}                   & not described\\
\rowcolor[gray]{.9} \citet{KangurKleis:1988}                    & not described by \citet{KimDunn:2007}\\
                    \citet{SondergaardEtal:1990}                & 'precision ball bearings'$^{(3)}$\\
\rowcolor[gray]{.9} \citet{FoersterEtal:1994}                   & G200$^{(4)}$, ashed, or machine finish\\
                    \citet{LorenzEtal:1997}                     & 'affordable', deviation from sphericity up to 50 \mum\\
\rowcolor[gray]{.9} \citet{LabousEtal:1997}                     & not described\\
                    \citet{FalconEtal:1998}                     & $\pm 1$ \mum\ tolerance, $\pm 0.8$ \mum\ sphericity$^{(5)}$\\
\rowcolor[gray]{.9} \citet{GorhamKharaz:2000}                   & 'deviation from sphericity of $< 0.12$ \mum'$^{(5)}$\\
                    \citet{LougeAdams:2002}                     & G10 or G25$^{(4)}$ (expected though not clearly described in the paper)\\
\rowcolor[gray]{.9} \citet{WeirTallon:2005}                     & not described\\
                    \citet{StevensHrenya:2005} & not described\\
\rowcolor[gray]{.9} \citet{ImreEtal:2008}                       & target polished\\
                  \citet{GrasselliEtal:2009} & not described \\
\rowcolor[gray]{.9} \citet{SoraceEtal:2009}                     & not described\\
                    \citet{DurdaEtal:2011a}                     & 'smooth but not polished'\\ 
\rowcolor[gray]{.9} \citet{KingEtal:2011} & better than 0.5 \mum, see their Fig. 3\\
                    \citet{MontaineEtal:2011} & tiny scratches \\
\rowcolor[gray]{.9} \citet{Beger_dipl}                          & within $\pm 0.1$ \mum\ (with few 1 \mum\ asperities; 'smooth') or $\pm 5$ \mum\ ('etched')\\
\\[-0.9em] \hline \\ 
\multicolumn{2}{c}{\textsc{Experiments with ice particles}}\\ \hline
                    reference                                   & surface condition\\ \hline\\[-0.9em]
\rowcolor[gray]{.9} \citet{BridgesEtal:1984, BridgesEtal:1996}, & smooth, rough through sublimation, or frost-covered\\	
\rowcolor[gray]{.9} \citet{HatzesEtal:1988, HatzesEtal:1991}    & \\	
                    \citet{DilleyCrawford:1996}                 & frosted\\
\rowcolor[gray]{.9} \citet{HigaEtal:1996, HigaEtal:1998}        & smooth, polished, frosted\\
                    \citet{HeisselmannEtal:2010}                & slightly frosted (ice), glass beads $\pm 5$ \mum\ \citep[same as][]{Beger_dipl}\\
\\[-0.9em] \hline
\end{tabular}\end{center}
$^{(1)}$ The authors are from the same group as \citet{DunnEtal:1995}, who stated their SEM resolution as 40 \AA.\\
$^{(2)}$ \citet{DunnEtal:1995} also described a mirror finish achieved with 0.05 \mum\ colloidal silica particle abrasives and quantified it to 'virtually flawless to within the stated resolution of the instrument (40 \AA)'. Pits which are resolved under an optical microscope should be of the order of 0.1 -- 1 \mum.\\
$^{(3)}$ Precision ball bearings are generally referred to as ball bearings with grades G100$^{(4)}$ or better.\\
$^{(4)}$ The grade G of a sphere is an industry norm for geometric tolerances used for ball bearings -- lower values mean better precision. We are interested here in the surface roughnesses, which are 0.02 \mum\ (G5), 0.03 \mum\ (G10), 0.05 \mum\ (G25), 0.08 \mum\ (G50), 0.13 \mum\ (G100), and 0.20 \mum\ (G200).\\
$^{(5)}$ If those spheres are fabricated according to industry norm, a deviation of sphericity of 0.13 \mum\ or 1.2 \mum\ would correspond to G5 or G50, respectively.
\vspace*{0.35cm}
\end{table*}

We have seen in the last section that many experiments show maxima in the coefficient of restitution well below unity, which are not as narrow as predicted by the model curve of \citet{ThorntonNing:1998}. In this context, the experiments of \citet{LifshitzKolsky:1964} are noteworthy as they described that it is possible to increase the level of this maximum by improving the surface conditions. They describe that the coefficients of restitution were initially -- for a 'finely ground surface finish' -- very scattered and never exceeded 0.82. With two polishing steps, first 'very fine emery paper' and then 'very fine abrasives', they could increase the coefficient of restitution to 0.87 and 0.95, respectively, and also the scatter became much smaller. The final surface was a 'highly reflecting mirror finish', however, with some pits detected under a light microscope. This can be indicative for not enough abrasion with the rougher grinding material and thus a too fast or steep change of this. In any case, a mirror finish was also described by \citet{DunnEtal:1995}, who quantified this qualitative impression with SEM analysis as flawless down to 40 \AA\ (steel). Although this is already a very good quality, \citeauthor{LifshitzKolsky:1964} argued that the coefficient of restitution might even become higher with a further improvement of the surface.

If we follow this line of arguments, we might conclude that there is a relation between surface quality of the samples and the value of the coefficient of restitution. This is not easy to establish from the given data but there are some indications. First, we compiled the surface conditions of the experiments provided in Table \ref{tab:overview_table} in a second listing, Table \ref{tab:surface_table}. For those publications where any information is made on the surface conditions, this is noted in the second column of Table \ref{tab:surface_table}. Some authors have explicitly measured the surface roughnesses of their projectiles and targets used. Methods contain SEM analysis, AFM measurements, and laser displacement analysis. In other experiments, ball bearings have been used with quality according to a certain industrial standard, mostly the one of the American Bearing Manufacturer Association (ABMA). If ball bearings were used but grade not stated in the publication, it can sometimes be concluded from other geometry tolerances given. It would be preferably to reduce the surface roughness to a single parameter like the arithmetic average deviation from a line profile
\begin{equation}
	R_\mathrm{a} = \frac{1}{n} \sum_{i=0}^n \left| y_i-\hat{y} \right|\ ,
\end{equation}
where $y_i-\hat{y}$ is the height difference of an $i$th point on a linear profile to the mean value $\hat{y}$. The definition for the roughness of a measured surface instead of a line works analog to this one and also a root-mean-squared roughness is sometimes used. Although all the information is reduced to a single value, the $R_\mathrm{a}$ value has the advantage that it is in principle easily accessible, either by manufacturer information or measurement.

We tried to provide these values in Table \ref{tab:surface_table} wherever possible. As some are given for surface measurements and others for profile lines, they are not exactly comparable, but it should be sufficient for a rough comparison. At this point however, we will leave this discussion to \citet{KrijtEtal:preprint_b} who do a more thoroughly analysis of the coefficients of restitution and relate it to their surface conditions.

Finally, it should be noted that also the cleanliness of the surfaces plays a decisive role. Dust floating around in the laboratory may have sizes of a micrometer and this is more than the surface roughnesses of most samples lieted in Table \ref{tab:surface_table}. Moreover, this dust is highly cohesive and samples must be carefully cleaned when exposed to this. Some authors describe a careful cleaning of their samples or targets \citep{WallEtal:1990, SondergaardEtal:1990, DunnEtal:1995, HigaEtal:1996, LiEtal:1999, PoppeEtal:2000a, GorhamKharaz:2000, Beger_dipl}.

\section*{Acknowledgements}
C.G. is grateful for the support of the Japan Society for the Promotion of Science (JSPS). D.H. acknowledges funding from the German Research Council (DFG) under grant Bl298/11-1.

{
\footnotesize
\bibliographystyle{apalike}
\balance
\bibliography{literature}
}

\end{document}